\documentclass[a4paper,fleqn]{cas-sc}

\def\tsc#1{\csdef{#1}{\textsc{\lowercase{#1}}\xspace}}
\tsc{WGM}
\tsc{QE}

\begin{document}
\let\WriteBookmarks\relax
\def\floatpagepagefraction{1}
\def\textpagefraction{.001}

\shorttitle{}    

\shortauthors{}  

\title [mode = title]{On the Consistency of GNN Explanations for Malware Detection}  



%

\author[1]{Hossein Shokouhinejad}

\cormark[1]


\ead{hossein.shokouhinejad@unb.ca}



\affiliation[1]{organization={Canadian Institute for Cybersecurity, University of New Brunswick},
            addressline={Sir Howard Douglas Hall}, 
            city={Fredericton},
            postcode={E3B 5A3}, 
            state={New Brunswick},
            country={Canada}}

\author[1]{Griffin Higgins}
\author[1]{Roozbeh Razavi-Far}
\author[1]{Hesamodin Mohammadian}
\author[1]{Ali A. Ghorbani}

\cortext[1]{Corresponding author}

\fntext[1]{}


\begin{abstract}
Control Flow Graphs (CFGs) are critical for analyzing program execution and characterizing malware behavior. With the growing adoption of Graph Neural Networks (GNNs), CFG-based representations have proven highly effective for malware detection. This study proposes a novel framework that dynamically constructs CFGs and embeds node features using a hybrid approach combining rule-based encoding and autoencoder-based embedding. A GNN-based classifier is then constructed to detect malicious behavior from the resulting graph representations.
To improve model interpretability, we apply state-of-the-art explainability techniques, including GNNExplainer, PGExplainer, and CaptumExplainer, the latter is utilized three attribution methods: Integrated Gradients, Guided Backpropagation, and Saliency. In addition, we introduce a novel aggregation method, called RankFusion, that integrates the outputs of the top-performing explainers to enhance the explanation quality. We also evaluate explanations using two subgraph extraction strategies, including the proposed Greedy Edge-wise Composition (GEC) method for improved structural coherence.
A comprehensive evaluation using accuracy, fidelity, and consistency metrics demonstrates the effectiveness of the proposed framework in terms of accurate identification of malware samples and generating reliable and interpretable explanations.
\end{abstract}




\maketitle

\section{Introduction}
In the past few years, malware attacks have escalated dramatically, highlighting the limitations of conventional malware detection methods, such as signature-based techniques. While these methods are quick and widely utilized, they fall short in identifying zero-day and sophisticated malware threats \cite{li2019machine}. Consequently, there has been a significant shift towards integrating machine learning (ML) strategies, noted for their enhanced detection capabilities, adaptability to emerging threats, and reduced false positives. Numerous studies have thus shifted focus towards leveraging ML \cite{frederick2022corpus, bensaoud2024cnn} for malware detection and classification.

As malware becomes more complex and developers employ advanced tactics to evade detection, the need for robust representation of malware samples has become evident. Recent research highlights the effectiveness of graph-based models in depicting malware behavior, which facilitates improved detection performance. These models provide a detailed view of a program's execution, helping analysts understand the program's logic, pinpoint vulnerabilities, and expose malicious activities, including hidden or scrambled code. Various graph structures, including Control Flow Graph (CFG)~\cite{MalGNE, sun2021effective}, Function Call Graph (FCG)~\cite{FCG_1, cai2021learning}, and Application Programming Interface (API) Call Graph (ACG)~\cite{API_Call_Graph_1, DMalNet}, have been increasingly utilized to feed data into ML models designed for malware detection. Although initial efforts involved using these graph structures with CNN or RNN architectures, the advent of Graph Neural Networks (GNNs) with variants like Graph Convolutional Networks (GCNs)~\cite{GCN}, Graph Isomorphism Networks (GINs)~\cite{GIN}, GraphSAGE~\cite{GraphSAGE}, and Graph Attention Networks (GATs)~\cite{GAT} have demonstrated superior outcomes when integrated with CFGs and FCGs.

The complexity of graph-based models often results in a lack of clarity in their decision-making processes, which is particularly crucial in the field of malware detection. To address this, various explainability methods for GNNs have been developed to identify the key subgraphs that influence the model’s decisions~\cite{shokouhinejad2025recent}. Explainability techniques provide insights into how GNN models reach their decisions by highlighting the most important nodes, edges, and subgraphs that contribute to the prediction. These methods improve the interpretability of GNN models, making them more transparent and trustworthy, which is especially valuable in security-related applications such as malware detection.

A critical factor influencing the performance of explainers is the method used to generate the important subgraph based on edge importance weights. Most existing research employs a conventional approach that selects the top-weighted edges to construct the important subgraph. However, the quality of the generated subgraph directly affects the reliability of the explanation and the model's overall interpretability. Therefore, evaluating the effectiveness and correctness of explainers is essential to ensure consistent and meaningful insights.

To assess the quality of explainers, several quantitative metrics are introduced to complement qualitative analyses. Among these, fidelity is a widely recognized metric that measures how faithfully the explainer represents the model's behavior. Specifically, fidelity assesses whether the model's performance on the original graph is consistent with its performance on the identified important subgraph while showing a significant deviation when evaluated on the unimportant subgraph. Another useful metric is the explainer accuracy, which measures how well the important subgraph retains the crucial information needed for the model's decision. Ideally, the explainer accuracy should align closely with the target GNN test accuracy, confirming that the identified subgraph captures the essential decision-making components of the model. However, evaluating the explainer with these metrics without considering the robustness of these metrics can lead to misleading conclusions. For example, high fidelity might reflect overfitting to the original graph rather than true model understanding. Similarly, explainer accuracy may be influenced by noise in the data or the structure of the original graph, potentially compromising the validity of the evaluation. Therefore, ensuring the consistency and stability of these metrics under different graph configurations is crucial for a reliable assessment of the explainer's performance.

In this paper, we propose a graph-based malware detection framework that leverages dynamically generated CFGs and hybrid node embeddings derived from assembly instructions. A GNN model is trained to classify the resulting graphs as benign or malicious. To enhance interpretability, we incorporate multiple explanation techniques and introduce a novel aggregation-based approach to improve explanation quality. We further propose an effective subgraph extraction strategy and assess the framework’s robustness using fidelity and consistency-based evaluation metrics.

The key contributions of this article are as follows:
\begin{itemize}
\item A dynamic graph-based malware detection framework utilizing CFGs and a hybrid rule-based and autoencoder-based node embedding strategy.
\item A novel explanation aggregation method, RankFusion explainer, that enhances the informativeness and reliability of subgraph-based interpretations.
\item A new subgraph extraction technique, GEC, designed to generate well-connected, high-importance subgraphs.
\item A systematic evaluation using accuracy, fidelity, and a consistency metric that measures the sensitivity of explanations to input perturbations.
\end{itemize}

The structure of the paper is organized as follows: Section~\ref{sec:background} provides a review of the relevant background. Section~\ref{sec:framework} details the proposed framework, while Section~\ref{sec:exp} focuses on the explainability metrics in more detail. Section~\ref{sec:result}, presents the experimental results and analysis. Finally, Section~\ref{sec:conclusion} concludes the paper and outlines potential directions for future research.

\section{Background}\label{sec:background}
The growing complexity and sophistication of malware have driven the need for more advanced detection techniques capable of thoroughly analyzing and identifying malicious programs. In recent years, GNNs have emerged as a powerful tool for malware detection due to their ability to model complex relationships and dependencies within the structural data of malicious code \cite{shokouhinejad2025recent}. Unlike traditional machine learning models that rely on flat feature representations, GNNs leverage the inherent graph structure of malware, such as CFGs, to capture both local and global patterns. Recent advances in GNN-based malware detection have focused on improving scalability, interpretability, and adaptability to evolving threat landscapes. Moreover, explainability frameworks integrated with GNNs are becoming increasingly important, allowing security analysts to understand the rationale behind classification decisions, identify critical graph components, and uncover hidden attack patterns. In this section, we provide an overview of the state-of-the-art techniques in GNN-based malware detection and discuss their approaches.

Recent advancements in GNN-based malware detection have introduced innovative techniques to improve detection accuracy, scalability, and robustness. One notable work is the Spectral-based Directed Graph Network (SDGNet), which addresses the challenge of detecting malware using directed graphs. Traditional spectral-based GNNs struggle with directed graphs due to the asymmetry of adjacency matrices. SDGNet overcomes this by employing three weighted graph matrix normalization methods (normal, aggregation-based, and propagation-based) to transform directed graphs into symmetrical matrices. It then applies a Multi-aspect Directed GCN (MDGCN) to learn comprehensive graph representations from these matrices~\cite{SDGNet}.

A dynamic malware analysis approach is presented in the DMalNet framework \cite{DMalNet}, which constructs an API call graph from API call sequences and applies a hybrid feature encoder to extract semantic features from both API names and arguments using techniques like Word2Vec, feature hashing, and similarity encoding. A GNN combining a modified GIN and GAT is then used to learn both content and structural features from the graph, capturing complex relationships between API calls for effective malware detection and classification.

A robust ensemble model for malware detection is described in REMSF \cite{REMSF}, which leverages semantic feature fusion. The model extracts static and semantic features from PE files, including byte histograms, entropy, and string information. It constructs a heterogeneous graph to model the relationships between PE files, imported DLLs, and APIs. By combining different classifiers through ensemble learning, REMSF improves detection accuracy and captures complex semantic relationships more effectively.

Another noteworthy approach is MalwareExpert \cite{Guided}, an expert system that detects malicious binaries using a GNN-based model. It identifies essential functions in the analyzed sample and highlights the most critical subgraphs involved in malicious behavior, providing an explainable output to improve transparency and understanding of the detection process.

A few-shot malware classification model using a graph transformer with a triplet-loss function is introduced in \cite{Triplet-trained}. This method extracts CFGs from assembly-level code and applies a path-sampling algorithm to capture functional patterns. The graph transformer, equipped with an attention mechanism, selectively embeds attack pathways from the CFGs. The triplet-loss function enables the model to learn a disentangled feature space, improving classification even with limited samples.

Documentation-augmented malware detection is explored in DawnGNN \cite{DawnGNN}. It constructs API graphs from Windows API call sequences and enhances them with semantic information extracted from official Windows API documentation using a pre-trained BERT model. The enhanced API graphs are then processed using GAT, which learns contextual information and improves malware detection by capturing both structural and semantic relationships among API calls.

Temporal and structural feature learning for malware detection is introduced in TS-Mal \cite{TS-Mal}. This model extracts fine-grained temporal patterns from API call sequences using a TextRCNN-based temporal vector learning method. It then models the relationships between API categories using a heterogeneous graph and generates dense structural representations through GAT. By combining both temporal and structural features, TS-Mal enhances the model’s ability to detect complex malware patterns.

MalGNE \cite{MalGNE} presents a novel malware detection framework based on CFG node embedding in a low-dimensional space. It addresses limitations in node feature extraction by applying a unique instruction encoding rule to handle out-of-vocabulary (OOV) issues and reduce redundancy. The model processes node vectors through an aggregation layer and a sequence layer to extract execution sequence and aggregation features. These vectors are mapped into a low-dimensional continuous space, improving both detection accuracy and efficiency through GNN-based learning.

Recent advancements in malware detection have not only focused on improving accuracy but also on enhancing model interpretability using GNN explainers. GNN explainers aim to provide insights into how a model reaches its decisions by identifying the most influential graph components.
A widely adopted method in GNN explainability is GNNExplainer \cite{GNNExplainer}, which provides model-agnostic explanations for predictions made by any GNN model on tasks such as node classification, link prediction, and graph classification. GNNExplainer identifies a compact subgraph structure and a small subset of node features that are most influential for a GNN’s prediction. It formulates the explanation process as an optimization problem that maximizes the mutual information between a GNN’s prediction and the distribution of possible subgraph structures. By learning both a structural and feature mask, GNNExplainer generates clear and consistent explanations, enhancing the interpretability of complex GNN models.

Building on this foundation, SubgraphX \cite{SubgraphX} introduces a GNN explanation method that explains GNN predictions through subgraph exploration. Unlike methods that focus on individual nodes or edges, SubgraphX identifies important subgraphs using a Monte Carlo tree search (MCTS) algorithm. It evaluates the importance of subgraphs by computing Shapley values, which measure the marginal contribution of each subgraph to the model’s prediction. This approach allows SubgraphX to highlight key subgraph structures responsible for the classification decision, providing more intuitive and interpretable explanations.

In the context of malware classification, CFGExplainer \cite{CFGExplainer} is specifically designed to interpret GNN-based malware classification results from CFGs. CFGExplainer identifies the most influential subgraphs contributing to classification and provides insight into the importance of individual nodes (basic blocks) within these subgraphs. The framework employs a two-stage process: an initial learning stage where a deep learning model assigns importance scores to node embeddings, and an interpretation stage where these scores are used to prune the graph and identify critical subgraphs. CFGExplainer enhances interpretability by revealing both the structural and functional aspects of malware behavior.

To improve scalability and generalization, PGExplainer \cite{PGExplainer} introduces a parameterized explanation method for GNNs that provides consistent and generalized explanations for multiple instances. Unlike GNNExplainer, which generates explanations independently for each instance, PGExplainer employs a deep neural network to parameterize the generation of explanations, enabling it to generalize across different instances. PGExplainer models the explanatory subgraph as an edge distribution and generates explanations by optimizing the mutual information between the subgraph structure and the GNN’s prediction. This approach improves both scalability and efficiency, making it suitable for inductive settings where new instances can be explained without retraining the explainer.

Recent efforts have also explored advanced reinforcement learning strategies to improve intrusion detection in complex environments such as the Industrial Internet of Things (IIoT). Notable among these are heuristic and open-set intrusion detection systems leveraging Deep Q-Networks (DQN) and its variants for handling zero-day attacks and unknown threats with limited labeled data~\cite{Deep_Q1}. Additionally, real-time defense models such as RT-A3C enhance responsiveness in edge-enabled IIoT scenarios~\cite{RT}, while positive sampling techniques in self-supervised graph contrastive learning offer promising directions for robust graph-based representations~\cite{Select_Your}.

\section{Proposed Method}\label{sec:framework}

\begin{figure*}[h]
    \centering
    \includegraphics[width=\linewidth]{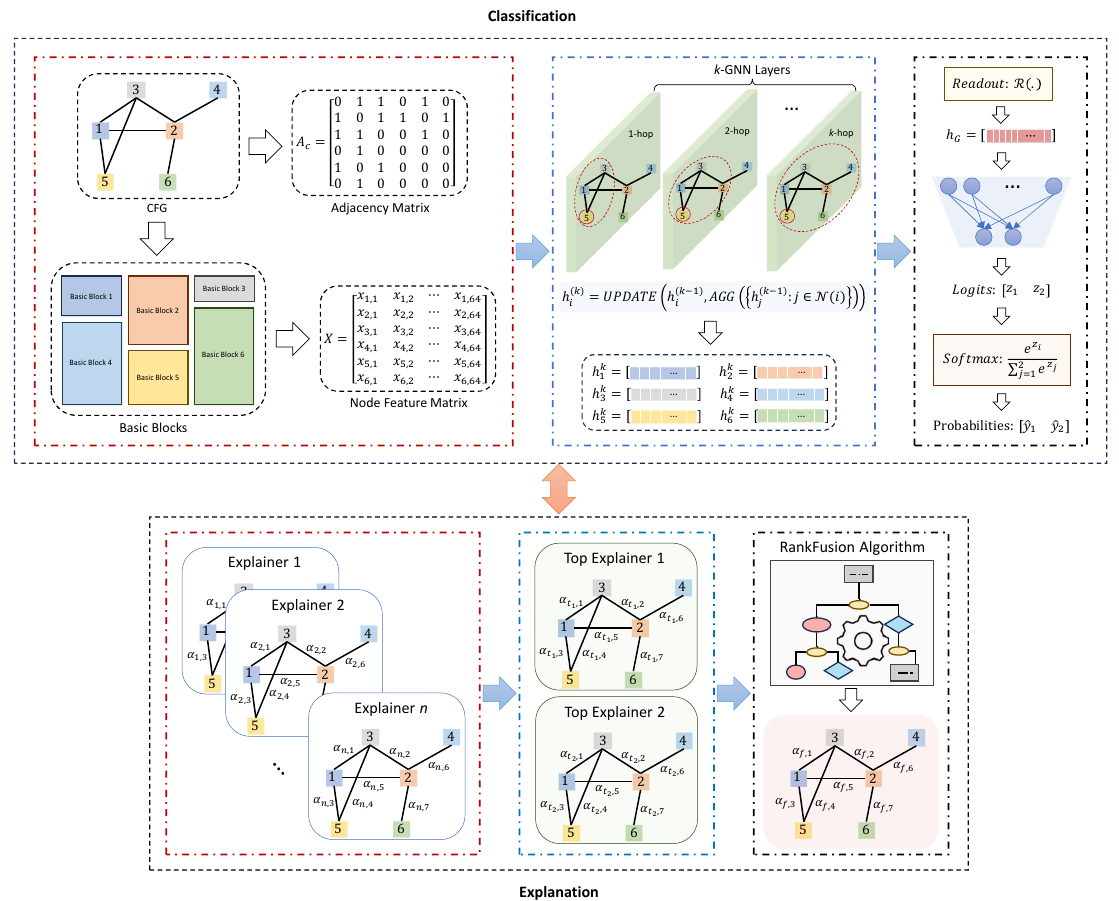}
    \caption{Proposed framework for interpretable malware detection.}
    \label{fig:Framework}
\end{figure*}
Our proposed framework for malware detection begins with dynamically generating a graph-based representation of the input sample through dynamic analysis. Specifically, a CFG is constructed, where each node represents a basic block of assembly instructions and edges represent control flow relationships between them. To enhance the accuracy and robustness of the detection model, a hybrid node feature embedding technique is used, which combines rule-based encoding and autoencoder-based embedding to convert the assembly instructions in each node into numerical vectors. A GNN-based model is then employed to classify the malware sample based on the graph structure and node features. To improve the interpretability of the model, the framework integrates GNNExplainer, PGExplainer, and CaptumExplainer to identify key subgraphs. Additionally, an aggregation scheme combines the edge rankings from the two top explainers to enhance the consistency of the extracted subgraphs. A new subgraph extraction method, GEC, is also introduced to generate more connected and meaningful subgraphs by prioritizing edges with the highest importance weights.
The overall structure of the proposed framework is illustrated in Figure \ref{fig:Framework}.

\subsection{Node Feature Embedding}
\begin{figure}[h]
    \centering
    \includegraphics[width=\linewidth]{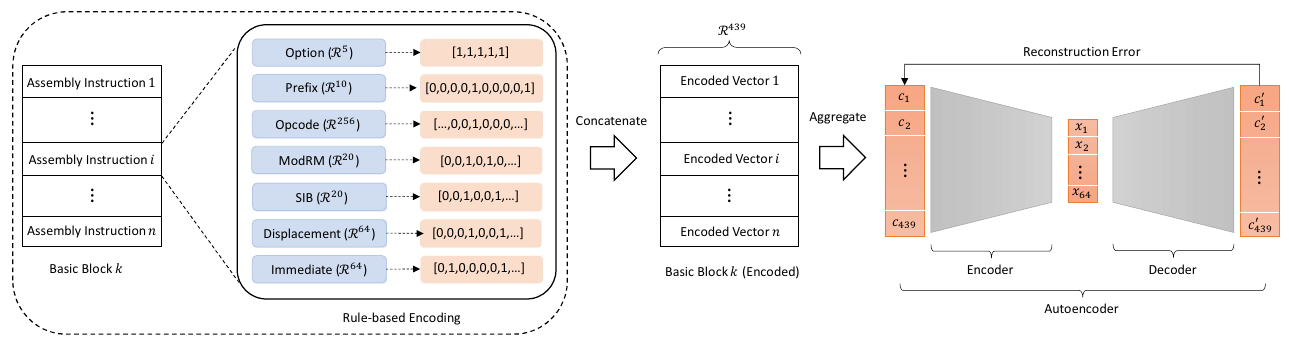}
    \caption{Embedding process for assembly instructions.}
    \label{fig:ae}
\end{figure}

The dynamically generated CFGs contain several types of features in each node (basic block), but not all are suitable or useful for the decision-making module. The assembly instructions of each node in the CFG are selected as the key features, and a two-step embedding technique is used to map them into a 64-dimensional real-valued vector. This process includes a rule-based instruction encoding strategy followed by machine learning-based dimensionality reduction using an autoencoder, as shown in Figure~\ref{fig:ae}.

\subsubsection{Rule-Based Instruction Encoding}
To encode the assembly instructions of CFG nodes, we adapted the encoding process outlined in~\cite{MalGNE} with slight modifications to better suit the structure of dynamically generated CFGs. Our approach applies a fine-grained rule-based encoding scheme that provides a unique and full-coverage mapping for the x86-64 instruction set, ensuring robust handling of all possible instructions. This design inherently addresses the out-of-vocabulary problem by mapping any instruction to a fixed 439-dimensional vector. Each x86-64 assembly instruction consists of up to seven components: option, prefix, opcode, ModRM, SIB, displacement, and immediate. The encoding process for each component is defined as follows:
\begin{itemize}
    \item \textbf{Prefix:} The prefix includes four fields: the extra segment (ES) register, operand-size override, address-size override, and lock prefix. The ES segment register has seven possible values, while the other three fields are binary (0 or 1). Thus, the prefix is encoded as a ten-dimensional one-hot vector.
    \item \textbf{Opcode:} The opcode defines the core operation performed by the instruction, with 256 possible values. It is encoded as a 256-dimensional one-hot vector.
    \item \textbf{ModRM:} ModRM is a one-byte field divided into three segments: two bits for the mode field, three bits for the register field, and three bits for the memory address field. This corresponds to four, eight, and eight possible values, respectively. Therefore, ModRM is encoded as a 20-dimensional one-hot vector.
    \item \textbf{SIB:} SIB (Scale-Index-Base) is another one-byte field divided into three segments: two bits for the scale factor, three bits for the index, and three bits for the base register. This results in four, eight, and eight possible values, respectively, leading to a 20-dimensional one-hot vector for SIB encoding.
    \item \textbf{Displacement:} In x86-64 assembly, displacement is an offset value used in memory addressing to calculate the effective memory address. It is part of the instruction's addressing mode and is added to a base or index register to determine the actual memory address being accessed. It is represented using a 64-dimensional binary vector.
    \item \textbf{Immediate} An immediate value is a constant value that is directly encoded as part of the instruction itself. Unlike a displacement, which is used for memory addressing, an immediate is used as an operand directly in the instruction. It is represented as a binary vector with 64 dimensions.
    \item \textbf{Option:} Since prefix, ModRM, SIB, displacement, and immediate fields are optional, a 5-dimensional binary vector is used to indicate their presence or absence.
\end{itemize}

The final encoded vector for each instruction is formed by concatenating all these components, resulting in a 439-dimensional vector.

Since a single CFG node may contain multiple instructions, the instruction vectors within a node are aggregated using a general aggregation function, which can be mean pooling, max pooling, or a learnable attention-based mechanism:  
\begin{equation}
\mathbf{E}_{\text{node}} = \text{AGG}(\mathbf{E}_{\text{instr}}^{(1)}, \mathbf{E}_{\text{instr}}^{(2)}, \ldots, \mathbf{E}_{\text{instr}}^{(n)})
\end{equation}
where $n$ is the number of instructions within the node, $\text{AGG}(.)$ represents the aggregation function (which can be mean, max, or attention-based), and $\mathbf{E}_{\text{node}} \in \mathbb{R}^{439}$ is the final aggregated vector for the node.

\subsubsection*{Dimensionality Reduction}  
The high-dimensional 439-dimensional vectors are reduced to a 64-dimensional latent representation using an autoencoder. The autoencoder consists of an encoder-decoder architecture, where the encoder reduces the dimensionality, and the decoder reconstructs the original vector. The reduced representation retains the most relevant information for malware detection.  

The autoencoder is optimized using a mean squared error (MSE) loss function:  
\begin{equation}
L_{\text{MSE}} = \frac{1}{N} \sum_{i=1}^{N} \| \mathbf{E}_{\text{instr}}^{(i)} - g_{\phi}(f_{\theta}(\mathbf{E}_{\text{instr}}^{(i)})) \|^2
\end{equation}
where \( N \) is the number of training samples, \( f_{\theta} \) is the encoder function, and \( g_{\phi} \) is the decoder function. Once trained, the encoder reduces the 439-dimensional instruction vector into a compact 64-dimensional feature vector used as the final node feature embedding for the GNN model:  
\begin{equation}
\mathbf{E}_{\text{node}}' = f_{\theta}(\mathbf{E}_{\text{node}})
\end{equation}
where \( \mathbf{E}_{\text{node}}' \in \mathbb{R}^{64} \).

This architecture offers key advantages in our context. Since only the entire CFG is labeled and individual basic blocks (nodes) do not have labels, end-to-end training approaches such as a Multi-Layer Perceptron (MLP) are not suitable. The autoencoder enables unsupervised dimensionality reduction, allowing the model to learn meaningful node representations without requiring node-level supervision. Additionally, attempting to jointly train a GNN with raw 439-dimensional input features and the graph structure may introduce instability during optimization and increase the risk of overfitting, particularly in security-sensitive domains. The autoencoder thus serves as an effective intermediate step that balances expressive representation with training stability.

\subsection{Graph Classification Using GNNs}  
After generating node feature embeddings, the next step is to classify the graph using a GNN model. The classification process involves three key stages: node embedding through GNN layers, graph-level representation generation using a readout layer, and final classification using a downstream classifier.  

\subsubsection{Node Embedding with GNN Layers}
A GNN model generates node embeddings by iteratively aggregating information from neighboring nodes through $k$ layers~\cite{NCP}. The node embedding at layer $l$ is computed using the following general update rule:  
\begin{equation}
h^{(l)}_i = \text{UPDATE}\left(h^{(l-1)}_i, \text{AGG}\left(\{h^{(l-1)}_j : j \in \mathcal{N}(i)\}\right)\right)
\end{equation}
where $h^{(l)}_i$ is the embedding of node $i$ at layer $l$, $h^{(0)}_i$ represents the initial node feature vector, $\mathcal{N}(i)$ is the set of neighboring nodes of node $i$, and $\text{UPDATE}(\cdot)$ is an update function that combines the aggregated information with the node’s current embedding.  

The aggregation function $\text{AGG}(\cdot)$ and update function $\text{UPDATE}(\cdot)$ can vary depending on the specific GNN model. For instance, in the case of GCNs, the aggregation is based on a normalized summation of neighboring node features, and the update step applies a linear transformation followed by a non-linearity:
\begin{equation}
    h^{(l)}_i = \sigma\left(\sum_{j \in \mathcal{N}(i) \cup \{i\}} \frac{1}{\sqrt{d_i d_j}} W^{(l)} h^{(l-1)}_j\right)
\end{equation}
where $\mathcal{N}(i)$ denotes the set of neighboring nodes of node $i$, $\sigma(\cdot)$ is a non-linear activation function such as ReLU, $W^{(l)}$ is the learnable weight matrix at layer $l$, and $d_i$ and $d_j$ denote the degrees of nodes $i$ and $j$, respectively. The term $\frac{1}{\sqrt{d_i d_j}}$ represents a normalization factor that accounts for the degree of each node to ensure stability during training.

This message-passing mechanism allows the node embeddings to incorporate information from neighboring nodes within a $k$-hop neighborhood, enabling the model to capture local graph structure and node attribute interactions.  

\subsubsection{Readout Layer}  
Once the node embeddings are computed through $k$ GNN layers, a readout layer generates a fixed-size graph-level representation by aggregating the final node embeddings. A general readout function can be defined as:  
\begin{equation}
h_G = \mathcal{R}(\{h^{(k)}_i : i \in V\})
\end{equation}
where $h_G$ is the graph-level representation, $V$ is the set of nodes in the graph, and $\mathcal{R}(\cdot)$ represents the readout function. Common readout functions include Mean pooling, Sum pooling, Max pooling, Attention-based pooling, Set2Set, and Sort Pooling.

\begin{algorithm}[h]
\caption{Node Feature Embedding}
\label{alg:embedding_pipeline}
\begin{algorithmic}[1]
\State \textbf{Input:} Set of CFG nodes $\mathcal{N}$, each with a list of instructions $\mathcal{I}_n = \{I_1, I_2, \dots, I_m\}$
\State \textbf{Output:} Embedded node features $\{\mathbf{E}_{\text{node}}'^{(1)}, \mathbf{E}_{\text{node}}'^{(2)}, \ldots\}$

\For{each node $n \in \mathcal{N}$}
    \State Initialize list of instruction vectors $\mathcal{V}_n \gets []$
    \For{each instruction $I \in \mathcal{I}_n$}
        \State Extract instruction components: prefix, opcode, ModRM, SIB, displacement, immediate, option
        \State Encode each component into its corresponding one-hot or binary vector
        \State Concatenate all components into a 439-dimensional vector $\mathbf{v}_I$
        \State $\mathcal{V}_n \gets \mathcal{V}_n \cup \{\mathbf{v}_I\}$
    \EndFor
    \State Aggregate instruction vectors: $\mathbf{E}_{\text{node}} \gets \mathcal{AGG}(\mathcal{V}_n)$
    \State Reduce dimensionality: $\mathbf{E}_{\text{node}}' \gets f_{\theta}(\mathbf{E}_{\text{node}})$
\EndFor
\State \Return Embedded features $\{\mathbf{E}_{\text{node}}'^{(1)}, \mathbf{E}_{\text{node}}'^{(2)}, \ldots\}$
\end{algorithmic}
\end{algorithm}

The rule-based instruction encoding is used to ensure robustness, full instruction coverage, and avoidance of out-of-vocabulary issues by mapping any valid instruction to a fixed-length 439-dimensional vector. This vector representation is semantically expressive, preserving the key structure and meaning of x86-64 instructions. The aggregation step enables flexible handling of varying numbers of instructions in each node. The dimensionality reduction through a trained autoencoder compresses the high-dimensional features into compact 64-dimensional embeddings, while preserving the most relevant information for the downstream classification task. This step significantly reduces computational overhead for GNN models without compromising the expressiveness of node features. The overall node feature embedding process, from rule-based instruction encoding to dimensionality reduction via autoencoder, is summarized in Algorithm~\ref{alg:embedding_pipeline}.

\subsubsection{Graph Classification} 
The graph-level representation $h_G$ is passed to a classifier, which is typically a fully connected neural network (FCNN) followed by a softmax activation to produce class probabilities:  
\begin{equation}
\hat{y} = \text{softmax}(W_c h_G + b_c)
\end{equation}
where $\hat{y}$ is the predicted class probability vector, $W_c$ is the weight matrix for the classifier, and $b_c$ is the bias term for the classifier.  

This process enables the GNN model to learn a graph-level representation that captures both node features and structural information.

\subsection{GNN Explanation Techniques}
While GNNs have demonstrated strong performance in graph-based learning tasks, their decision-making processes often remain opaque. Explainability methods aim to interpret how GNN models arrive at their predictions by identifying the most influential nodes, edges, or subgraphs within the input graph. In this study, we employ five state-of-the-art explainers: GNNExplainer, PGExplainer, and CaptumExplainer with three attribution methods: Integrated Gradients, Guided Backpropagation, and Saliency. All these explainability techniques assign weights to the edges based on their importance to the decision made by the GNN model. After evaluating their performance, we select the two best-performing explainers and propose a new explainer based on an aggregation method that combines the edge rankings from these two explainers to generate more consistent and informative subgraphs.

\begin{figure}[h]
    \centering
    \includegraphics[width=0.9\linewidth]{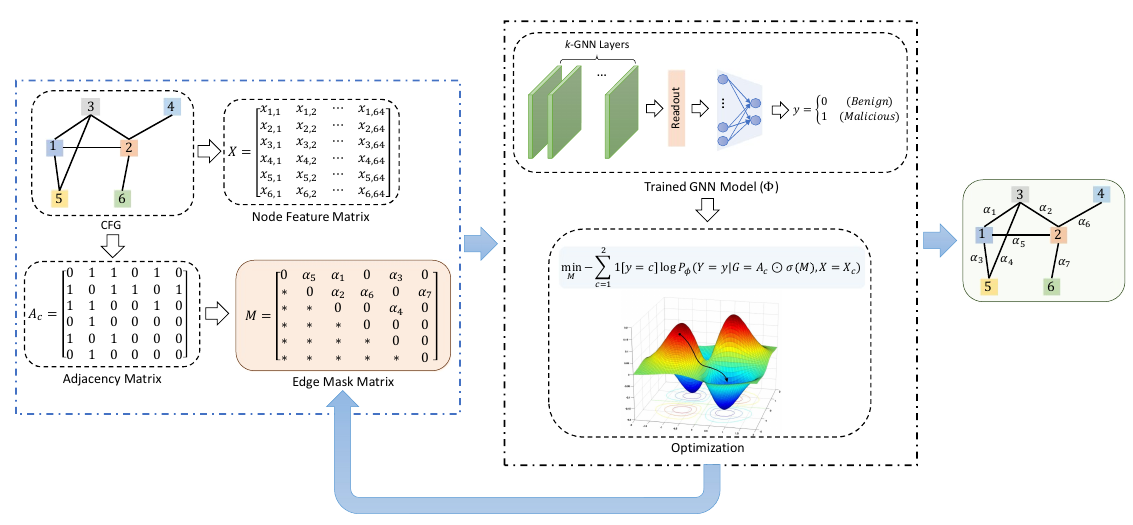}
    \caption{GNNExplainer generates explanations by optimizing masks per sample.}
    \label{fig:gnnexplainer}
\end{figure}

GNNExplainer and PGExplainer are based on the key idea of identifying a subgraph that maximizes the mutual information ($MI$) with the model's prediction. This objective can be formulated as:
\begin{equation}
\underset{G_{s}}{\max}\, MI\,(Y, (G_{s}, X_{s})) = H(Y) - H(Y|G = G_{s}, X = X_{s})
\end{equation}
where $G_s$ and $X_s$ represent the explanatory subgraph and its associated node features, respectively, $Y$ is the prediction of the GNN model, and $H(\cdot)$ denotes the entropy function. The goal is to find the most informative subgraphs and node features that contribute to the model's prediction.
Since $H(Y)$ is constant for a trained GNN, maximizing the mutual information reduces to minimizing the conditional entropy. GNNExplainer proposes approximating the conditional entropy using the cross-entropy loss between the true class label and the model's prediction, resulting in the following optimization objective for binary classification, which can be optimized using gradient descent:
\begin{equation}
    \underset{M}{\min}\,-\sum_{c=1}^{2}\mathds{1}[y=c]\,\text{log}\, P_{\Phi}(Y=y|G=A_{c}\odot\sigma(M),X=X_{c})
\end{equation}
where $c$ is the class label, $y$ is the predicted label, $A_{c}$ is the adjacency matrix of the input graph, $M$ is the learnable mask that assigns a weight to each edge of the input graph based on its importance for the model's prediction, $\sigma$ is the sigmoid function that maps the weight to the range $[0,1]$, and $X_c$ is the corresponding node feature set. The indicator function $\mathds{1}[y = c]$ equals $1$ if $y = c$, and $0$ otherwise. Moreover, $P_{\Phi}(\cdot)$ denotes the prediction probability distribution output by the trained GNN model $\Phi$ for a given input graph $G$ and node features $X$. This optimization is performed separately for each input sample to generate its corresponding explanation. GNNExplainer jointly learns an edge mask and a node feature mask to identify the most influential structural and attribute components of the graph. After optimization, the top-ranked edges based on their weights are selected to form the important subgraph. The overall process of GNNExplainer is illustrated in Figure~\ref{fig:gnnexplainer}.

\begin{figure}[h]
    \centering
    \includegraphics[width=\linewidth]{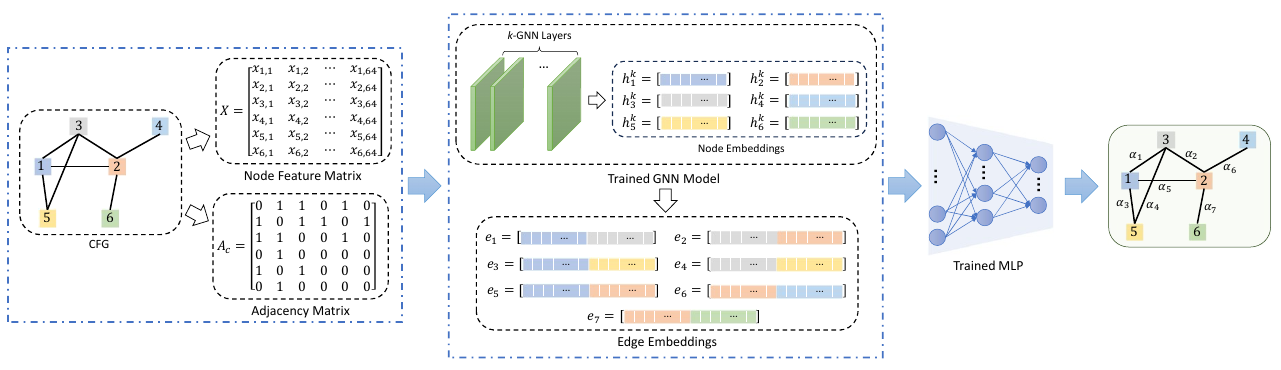}
    \caption{Global explanation via MLP-based edge scoring in PGExplainer.}
    \label{fig:pgexplainer}
\end{figure}

PGExplainer uses the following cross-entropy loss to train the explainer on multiple samples. This explainer acts as a global explainer by training a shared explanation model across multiple samples. Specifically, it learns an MLP that predicts edge importance scores. For binary classification, the objective is formulated as:
\begin{equation}
    \underset{\Psi}{\min}\,-\sum_{i\in I}\sum_{k=1}^{K}\sum_{c=1}^{2}P_{\Phi}(Y=c|G=G_{o}^{(i)})\,\text{log}\,P_{\Phi}(Y=c|G=\hat{G}_{s}^{(i,k)})
\end{equation}
where $G_{o}^{(i)}$ is the original input graph for sample $i$, $\hat{G}_{s}^{(i,k)}$ is the sampled explanatory subgraph for sample $i$ in the $k$-th sampling step, and $\Psi$ represents the parameters of the explainer model. The architecture of PGExplainer is depicted in Figure~\ref{fig:pgexplainer}.

In addition to GNNExplainer and PGExplainer, we employed CaptumExplainer to enhance the interpretability of GNN models. CaptumExplainer provides a framework for attributing the model’s predictions to individual nodes, edges, and features, allowing a detailed understanding of how the model processes graph data. We utilized three key attribution methods within CaptumExplainer: Integrated Gradients, Saliency, and Guided Backpropagation.

Integrated Gradients computes the contribution of each input feature by integrating the gradients of the model's output with respect to the input along a path from a baseline to the actual input. This method satisfies two key axioms: sensitivity, which ensures that attributions reflect the difference in the model's outputs between the baseline and the input, and implementation invariance, which guarantees consistent attributions across functionally equivalent models.

Saliency measures the influence of each input feature by calculating the gradient of the model's output with respect to the input, effectively performing a first-order Taylor expansion. The magnitude of these gradients reflects the importance of each feature in the model's decision.

Guided Backpropagation refines the saliency method by modifying the backpropagation process to propagate only non-negative gradients through ReLU activations, thereby highlighting the most impactful and positively contributing features. These three methods provide complementary insights into the model’s decision-making process, enabling a deeper understanding of the structural and feature-based factors that drive GNN predictions.

\subsection{RankFusion Explainer: An Edge Ranking Aggregation Strategy}
To enhance the quality of explanations across different sparsity levels, we propose a novel GNN explanation method called RankFusion explainer. This approach aggregates the edge rankings from two state-of-the-art explainers to generate more stable and accurate explanations. The process begins with constructing a bank of candidate explainers, including GNNExplainer, PGExplainer, CaptumExplainer, and other techniques. Each explainer in this bank is individually assessed by computing its accuracy across a range of sparsity levels. This is done by retaining the top-ranked edges, reconstructing the graph, and measuring the GNN’s prediction accuracy. Based on the overall performance, top two explainers are selected for fusion.

After the selection stage, the RankFusion algorithm transforms the edge importance scores from both explainers into descending rank orders. For each edge, the algorithm calculates the absolute difference between their assigned ranks. If this difference is smaller than a predefined threshold (expressed as a percentage of the total number of edges), the fused rank is conservatively assigned as the maximum of the two ranks. If the rank difference exceeds the threshold, the rank of that edge is taken from the explainer with higher accuracy at the corresponding explanation percentage, ensuring that disagreements are resolved in favor of the more reliable explainer.

Once all edges have been assigned fused rankings, the top-ranked edges based on the new weights are selected to reconstruct a sparsified version of the input graph. This reconstructed subgraph is then fed into the trained GNN model, and its prediction accuracy is recorded as the evaluation metric for the RankFusion explainer. This fusion-based strategy balances consensus and performance priority, leveraging the strengths of multiple explainers to improve robustness and fidelity in graph-based explanations across varying sparsity constraints. The detailed algorithmic steps of the RankFusion explainer are presented in Algorithm~\ref{alg:rankfusion}.

\begin{algorithm}[h]
\caption{RankFusion}
\label{alg:rankfusion}
\begin{algorithmic}[1]
\State \textbf{Input:} Graph $G = (V, E)$, 
    Edge weights from Explainer 1 ($\alpha_{t_1}$), 
    Edge weights from Explainer 2 ($\alpha_{t_2}$), 
    Accuracy of Explainer 1 ($A_1$) and Explainer 2 ($A_2$), 
    Threshold percentage ($T$)
\State \textbf{Output:} New edge weights ($\alpha_{\text{f}}$)

\State $n \gets |E|$ 
\State Threshold $\tau \gets T \cdot n / 100$  

\For{each edge $e$}  
    \State $d \gets |\alpha_{t_1}[e] - \alpha_{t_2}[e]|$ 
    \If{$d \leq \tau$}  
        \State $\alpha_{\text{f}}[e] \gets \max(\alpha_{t_1}[e], \alpha_{t_2}[e])$ 
    \Else  
        \If{$A_1 \geq A_2$}  
            \State $\alpha_{\text{f}}[e] \gets \alpha_{t_1}[e]$  
        \Else  
            \State $\alpha_{\text{f}}[e] \gets \alpha_{t_2}[e]$  
        \EndIf  
    \EndIf  
\EndFor  

\State \Return $\alpha_{\text{f}}$

\end{algorithmic}
\end{algorithm}

\section{Explainer Subgraph Evaluation and Construction}
\label{sec:exp}

An effective GNN explainer should identify a subgraph that preserves the model’s predictive capability while revealing the key structural features driving the prediction. Ideally, an important subgraph should produce a prediction outcome close to that of the original input. At the same time, the model's performance should drop significantly, when it processes the remaining part of the graph after the important subgraph is removed. This behavior confirms that the identified subgraph contains the critical features influencing the model's decision-making.
To assess the quality of explainer-generated subgraphs, we rely on two primary evaluation metrics: explainer accuracy and fidelity.

Explainer accuracy measures the model's performance, when using only the subgraph identified by the explainer as input. A well-constructed subgraph explanation should preserve the model's original classification performance and accuracy. This metric operates under the assumption that the identified subgraph retains the most influential features necessary for accurate predictions.

Fidelity measures the contribution of the important subgraph to the model's prediction. It evaluates how much the prediction changes, when the important or unimportant parts of the graph are removed. Fidelity is defined two complementary forms:

\begin{equation}
    Fidelity{+} = \frac{1}{N}\sum_{i=1}^{N}\left ( f\left ( G^{(i)} \right )_{y_i} - f\left ( G^{(i)}-G_{s}^{(i)} \right )_{y_i} \right ),
\end{equation}

\begin{equation}
    Fidelity{-} = \frac{1}{N}\sum_{i=1}^{N}\left ( f\left ( G^{(i)} \right )_{y_i} - f\left ( G_{s}^{(i)} \right )_{y_i} \right ),
\end{equation}
where $G_s^{(i)}$ is the important subgraph, $G^{(i)}$ is the original input, and $f(.)$ is the trained GNN model and $y_i$ is the ground truth label. $Fidelity{+}$ measures the prediction difference between the original input and its unimportant part, capturing the contribution of the removed edges to the model’s decision. Conversely, $Fidelity{-}$ compares the prediction between the original graph and the important subgraph, reflecting the predictive capacity of the extracted subgraph. A higher $Fidelity{+}$ value indicates that removing the identified important subgraph significantly alters the model’s prediction, confirming that this subgraph is crucial for the decision made. Therefore, higher $Fidelity{+}$ scores correspond to more faithful explanations.
In contrast, a lower $Fidelity{-}$ value means that the important subgraph alone can nearly reproduce the model’s prediction on the original graph, suggesting it contains most of the essential information. Thus, lower $Fidelity{-}$ values reflect higher-quality explanations.

In addition to accuracy and fidelity, consistency is an essential property for evaluating the reliability of explainers. It reflects the stability of an explanation, when the input graph is subjected to small structural perturbations that do not alter the model’s prediction~\cite{hajiramezanali2023on}. A consistent explainer should yield similar explanations for slightly modified graphs, thereby enhancing interpretability and trustworthiness.

Let \( G = (V, E) \) be an input graph, with \( \mathcal{M}(G) \) denoting the prediction of a trained GNN model. To evaluate consistency, we generate a set of perturbed graphs \( \mathcal{G}_{\text{pert}} = \{G'_1, G'_2, \ldots, G'_m\} \), where each \( G'_i \) is created by randomly removing a small subset of nodes or edges from \( G \).

The number of removed elements is computed as:
\begin{equation}
    |S| = \left\lceil \log(|X|) + |X| \cdot p \right\rceil,
\end{equation}
where \( X \in \{V, E\} \), \( S \subseteq X \), and \( p \in (0, 1) \) is a perturbation ratio.

Each perturbed graph \( G'_i \) is retained for consistency evaluation if it meets the following two conditions. First, the model prediction must be unchanged:
\begin{equation}
    f(G'_i) = f(G)
\end{equation}

Second, the graph-level embeddings must remain similar:
\begin{equation}
    1 - \cos\left( \phi(G), \phi(G'_i) \right) < \tau,
\end{equation}
where \( \phi(\cdot) \) denotes the embedding function of the model, and \( \tau \) is a threshold for the similarity metric.

For each valid \( G'_i \), we compute the explainer’s fidelity scores $Fidelity{+}(G'_i)$ and $Fidelity{-}(G'_i)$. The consistency of the explainer is measured by the variability of these scores:
\begin{equation}
    \Delta^+ = \max_i Fidelity{+}(G'_i) - \min_i Fidelity{+}(G'_i)
\end{equation}
\begin{equation}
    \Delta^- = \max_i Fidelity{-}(G'_i) - \min_i Fidelity{-}(G'_i)
\end{equation}

Smaller values of \( \Delta^+ \) and \( \Delta^- \) indicate greater consistency, as they reflect less variation in the explanation quality under small, non-disruptive perturbations.

\subsubsection{Top-Edge Selection (TES)} Conventional methods for subgraph extraction rank edges according to their weights and select the highest-ranked edges to construct a subgraph of a specific size relative to the original graph. We refer to this approach as TES and use it as our baseline method.

However, extracting subgraphs using TES often results in explanations composed of multiple disconnected components, whereas original graphs typically consist of a single connected component. This discrepancy can reduce the reliability of evaluation metrics such as accuracy and fidelity, since the target GNN is trained on connected graphs and may not perform well on fragmented subgraphs.

\subsubsection{Greedy Edge-wise Composition (GEC)}

To address the limitations of TES, we propose Greedy Edge-wise Composition (GEC), an alternative subgraph extraction technique designed to construct strongly connected subgraphs with high cumulative edge importance. This method operates on a weighted graph representation \( G' = (V', E', \mathbf{S}_E)\), where \( \mathbf{S}_E \) is the edge importance scores.

The process begins by selecting the edge with the highest importance weight:
\begin{equation}
    e_{\text{max}} = \arg\max_{e \in E'} \alpha_e,
\end{equation}
where \( \alpha_e \in \mathbf{S}_E \) denotes the importance weight of edge \( e \). The two nodes incident to \( e_{\text{max}} \) are added to the selected node set \( V_{\text{selected}} \), and the edge itself is added to the selected edge set \( E_{\text{selected}} \).

GEC then proceeds iteratively, at each step selecting the next highest-weight edge that connects to the current set of selected nodes:
\begin{equation}
    e_{\text{next}} = \arg\max_{\substack{e \in E'}} \alpha_e,  \quad e \text{ connects to } V_{\text{selected}}
\end{equation}

The corresponding node(s) from \( e_{\text{next}} \) not already in \( V_{\text{selected}} \) are added to the node set, and \( e_{\text{next}} \) is added to the edge set. This greedy procedure is repeated until a predefined number of edges is included.

Let \( k \) be the target number of edges to select. The final extracted subgraph is:
\begin{equation}
    G_{\text{extracted}} = (V_{\text{selected}}, E_{\text{selected}}), \quad \text{where} \quad |E_{\text{selected}}| = k.
\end{equation}

Algorithm \ref{alg:gec} presents the GEC algorithm. By prioritizing both edge weight and structural connectivity, GEC ensures that the selected subgraph retains the most informative and influential components of the original graph. This results in more faithful and robust explanations, ultimately supporting a more reliable interpretation of the model’s decision-making process.
\begin{algorithm}[h]
\caption{Greedy Edge-wise Composition (GEC)}
\label{alg:gec}
\begin{algorithmic}[1]
\State \textbf{Input:} Graph $G' = (V', E', \mathbf{S}_E)$ with edge importance scores, target number of edges $k$
\State \textbf{Output:} Extracted subgraph $G_{\text{extracted}} = (V_{\text{selected}}, E_{\text{selected}})$
\vspace{1mm}

\State \textbf{Step 1: Initialization}
\State Sort all edges by descending importance into list $\mathcal{L}$
\State $E_{\text{selected}} \gets \emptyset$ 
\State $V_{\text{selected}} \gets \emptyset$ 
\State $used \gets$ array of False flags for all edges in $\mathcal{L}$
\vspace{1mm}

\State \textbf{Step 2: Add the most important edge}
\State Select the top edge $e_{\text{max}} \gets \mathcal{L}[0]$
\State Add $e_{\text{max}}$ to $E_{\text{selected}}$
\State Add both endpoint nodes of $e_{\text{max}}$ to $V_{\text{selected}}$
\State Mark $e_{\text{max}}$ as used
\vspace{1mm}

\State \textbf{Step 3: Grow subgraph iteratively}
\While{$|E_{\text{selected}}| < k$}
    \For{each unused edge $e$ in $\mathcal{L}$}
        \State Let $u$ and $v$ be the endpoint nodes of $e$
        \If{$u \in V_{\text{selected}}$ \textbf{or} $v \in V_{\text{selected}}$}
            \State Add $e$ to $E_{\text{selected}}$
            \State $V_{\text{selected}} \gets V_{\text{selected}} \cup \{u, v\}$
            \State Mark $e$ as used
            \State \textbf{break} 
        \EndIf
    \EndFor
\EndWhile
\vspace{1mm}

\State \Return $G_{\text{extracted}} = (V_{\text{selected}}, E_{\text{selected}})$
\end{algorithmic}
\end{algorithm}

\section{Results and Analysis}
For our experiments, we selected malicious samples from BODMAS~\cite{yang2021bodmas} and PMML~\cite{practicalsecurity2024pe}, along with benign samples from DikeDataset~\cite{dikedataset}. Details of the datasets used in this study are summarized in Table~\ref{tab:dataset_stats}.
To recover CFGs dynamically, we employ the Angr library~\cite{shoshitaishvili2016state}, a Python-based binary analysis tool. Angr constructs graphs by integrating both symbolic execution and constraint solving, enabling comprehensive and precise CFG recovery. The dynamically generated CFGs used in our experiments are available for public access\footnote{Available at \url{http://cicresearch.ca/CICDataset/CIC-DGG-2025/Dataset/}}. Moreover, the implementation used in this study has been made publicly available\footnote{Available at \url{https://github.com/GriffinUNB/On-the-Consistency-of-GNN-Explanations-for-Malware-Detection}}.

\begin{table}[h]
\centering
\caption{Characteristics of the evaluated datasets.}
\label{tab:dataset_stats}
\begin{tabular}{lcccc}
\hline
\textbf{Datasets} & \textbf{\#Samples} & \textbf{Avg. Nodes} & \textbf{Avg. Edges} & \textbf{Class} \\
\hline
BODMAS & 1117 & 57587.82 & 60252.08 & Malware \\
DikeDataset & 510 & 6011.56 & 10054.94 & Benign \\
PMML & 1029 & 11303.51 & 19073.31 & Malware \\
\hline
\end{tabular}
\end{table}
We employed a symmetrical autoencoder to reduce the original 439-dimensional node feature vectors to a 64-dimensional latent space. The encoder comprises three fully connected layers with dimensions 439~$\rightarrow$~256~$\rightarrow$~128~$\rightarrow$~64, each followed by a ReLU activation function. The decoder mirrors this structure with layers 64~$\rightarrow$~128~$\rightarrow$~256~$\rightarrow$~439, also using ReLU activations. The autoencoder was trained for 5000 epochs using the Adam optimizer with a learning rate of 0.0001, minimizing the mean squared error between the input and reconstructed features. Training was stopped at 5000 epochs because the validation MSE plateaued below a threshold of $1\times10^{-4}$ for the final 1000 epochs, indicating convergence. The 64-dimensional embeddings produced by the encoder were subsequently used as the input node features for downstream graph learning tasks.

\begin{figure}
    \centering
    \includegraphics[width=0.5\linewidth]{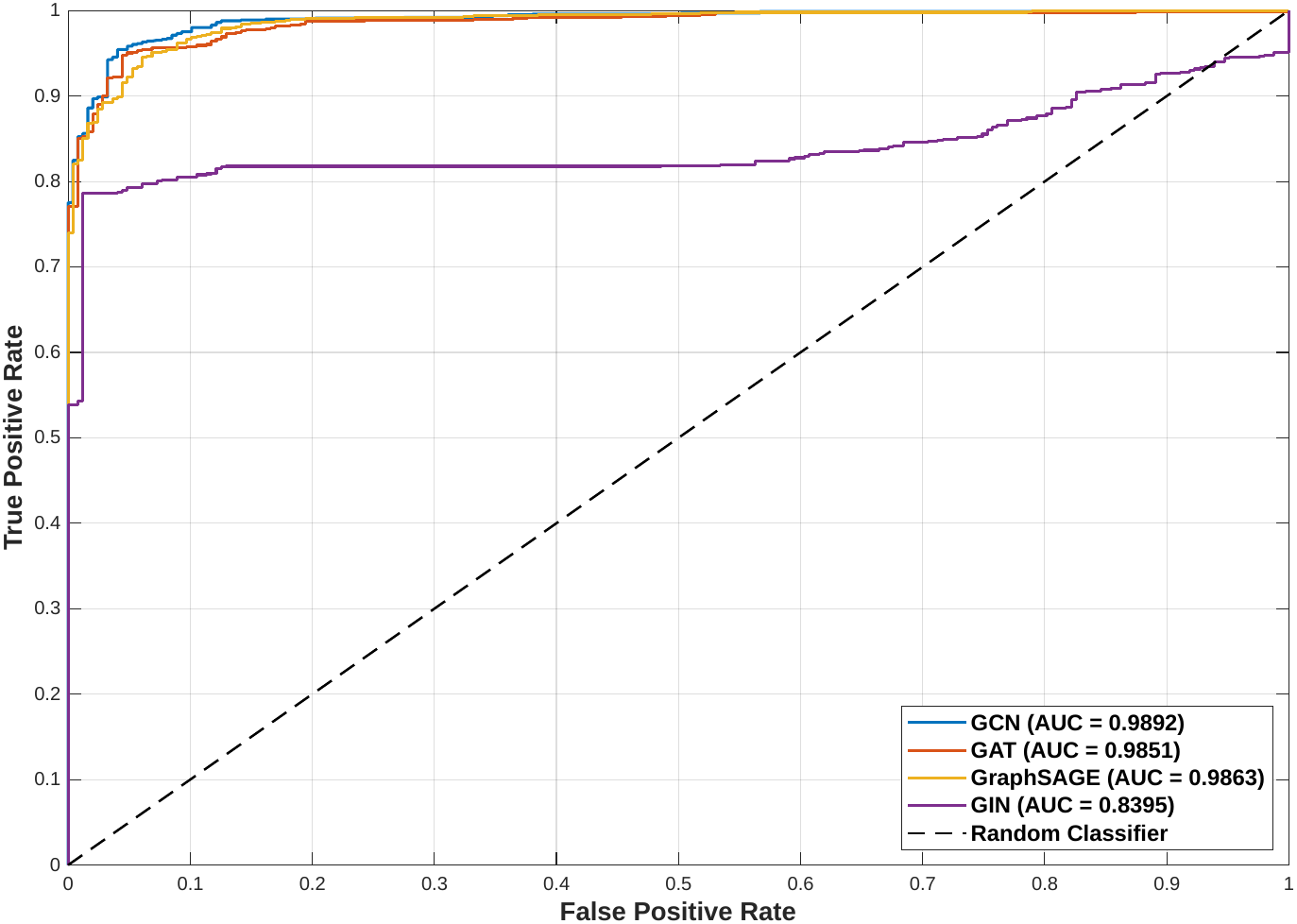}
    \caption{ROC curves for GNN models.}
    \label{fig:roc}
\end{figure}

For graph classification, we experimented with several GNN architectures, including GCN, GAT, GraphSAGE, and GIN. All models share the same architecture: three graph convolution layers with 64 hidden units each, followed by ReLU activations. A global mean pooling layer aggregates node-level embeddings into a graph-level representation, which is then passed through a dropout layer (with a dropout rate of 0.2) and a final fully connected linear layer to produce class scores for binary classification. Each model was trained using the Adam optimizer with a learning rate of 0.0001 and weight decay of 0.0005, using the cross-entropy loss over 50 epochs. The dataset was randomly split into 50\% for training and 50\% for testing. Table~\ref{tab:gnn_classwise_metrics} summarizes the class-wise evaluation metrics (precision, recall, and F1-score) for both benign and malicious classes, along with the overall accuracy for each model. Among the tested architectures, GCN outperformed the others, achieving the highest accuracy of 96.02\%, along with a benign F1-score of 0.8927 and a malicious F1-score of 0.9756. The ROC curves for each model are presented in Figure~\ref{fig:roc}, providing additional insight into their detection capabilities. Based on its superior overall performance, GCN was selected as the backbone model for the downstream explanation task.

\begin{table*}[h]
\centering
\caption{Class-wise evaluation metrics for GNN models on malware detection task.}
\label{tab:gnn_classwise_metrics}
\begin{tabular}{lccccccc}
\toprule
\multirow{2}{*}{\textbf{Model}} & \multirow{2}{*}{\textbf{Accuracy}} & 
\multicolumn{3}{c}{\textbf{Benign Class}} & 
\multicolumn{3}{c}{\textbf{Malicious Class}} \\
\cmidrule(lr){3-5} \cmidrule(lr){6-8}
& & \textbf{Precision} & \textbf{Recall} & \textbf{F1-score} & \textbf{Precision} & \textbf{Recall} & \textbf{F1-score} \\
\midrule
GCN~\cite{GCN}        & 0.9602 & 0.9498 & 0.8421 & 0.8927 & 0.9624 & 0.9891 & 0.9756 \\
GIN~\cite{GIN}        & 0.7056 & 0.4000 & 0.9960 & 0.5708 & 0.9984 & 0.6347 & 0.7760 \\
GraphSAGE~\cite{GraphSAGE}  & 0.9515 & 0.8577 & 0.9028 & 0.8797 & 0.9759 & 0.9634 & 0.9696 \\
GAT~\cite{GAT}        & 0.9523 & 0.8912 & 0.8623 & 0.8765 & 0.9666 & 0.9743 & 0.9704 \\
\bottomrule
\end{tabular}
\end{table*}

As previously discussed, the performance of the explainers is evaluated using the accuracy, fidelity, and consistency metrics. The remainder of this section presents a detailed comparison of five explainers: GNNExplainer, PGExplainer, and three variants of CaptumExplainer based on different attribution methods: Integrated Gradients, Guided Backpropagation, and Saliency. The evaluation is conducted through two different subgraph extraction techniques: TES and GEC, based on these three metrics. Additionally, our proposed RankFusion explainer is evaluated by selecting the top two best-performing explainers.

\begin{figure*}[h]
    \centering
    \begin{minipage}[t]{0.48\linewidth}
        \centering
        \includegraphics[width=\linewidth]{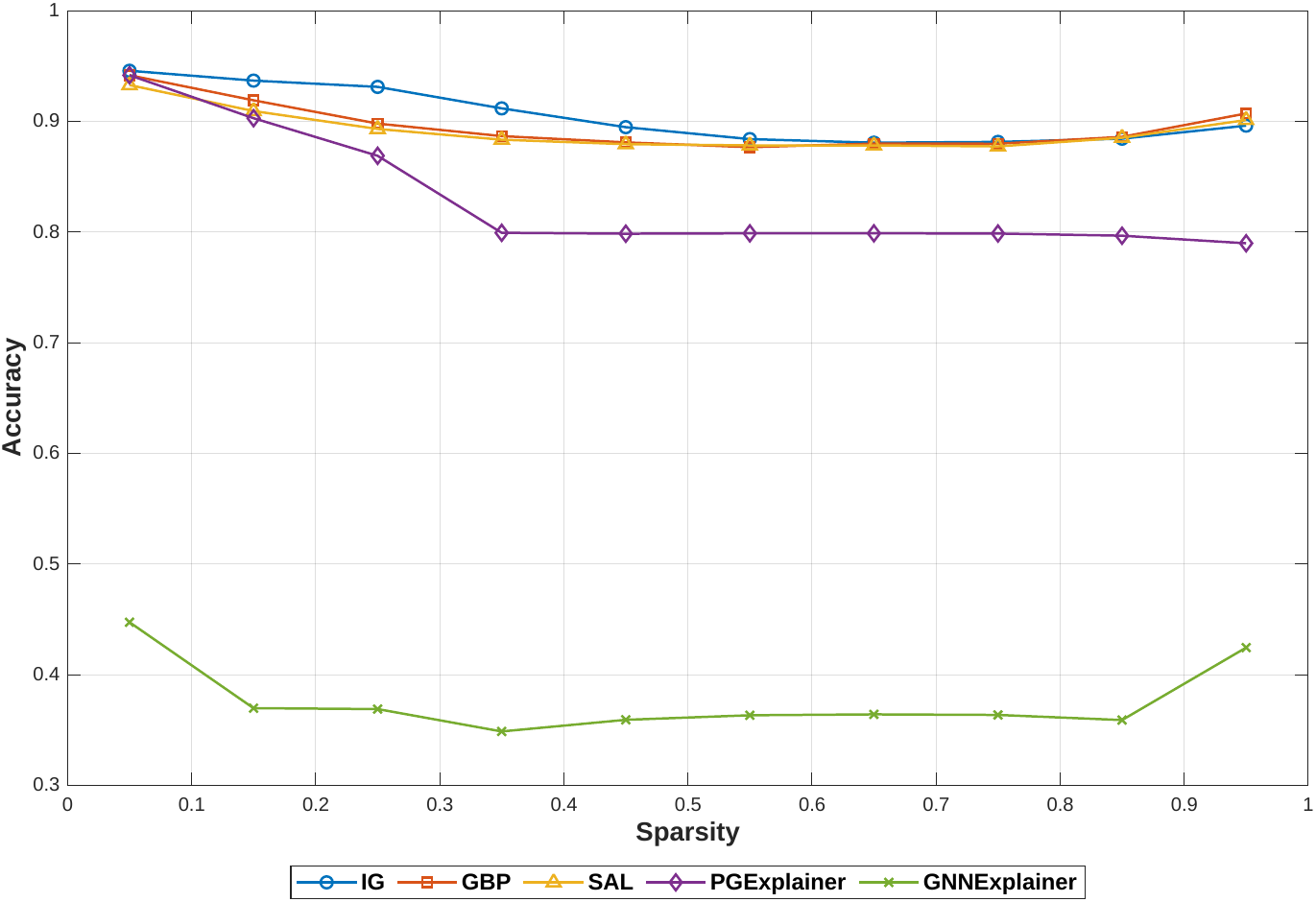}
        \subcaption{TES-based subgraph extraction.}
    \end{minipage}
    \hfill
    \begin{minipage}[t]{0.48\linewidth}
        \centering
        \includegraphics[width=\linewidth]{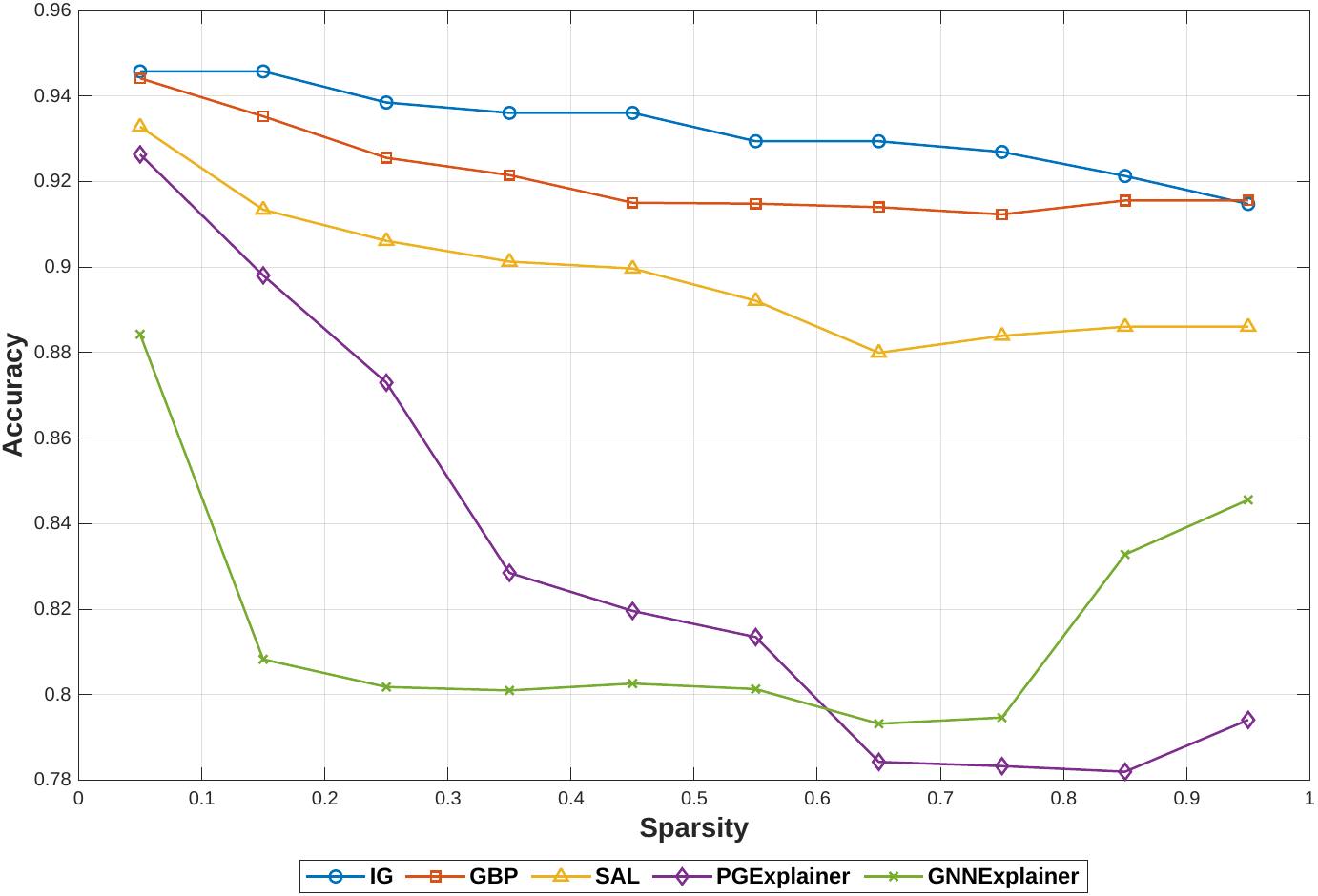}
        \subcaption{GEC-based subgraph extraction.}
    \end{minipage}
    \caption{Accuracy performance of explainability methods under TES and GEC subgraph extraction approaches.}
    \label{fig:acc_compare}
\end{figure*}

Figure~\ref{fig:acc_compare} presents a comparative analysis of model accuracy based on different explainability techniques and subgraph extraction methods. The x-axis denotes subgraph sparsity, defined as the percentage of edges retained from the original graph, ranging from 5\% to 95\%. The y-axis indicates the accuracy of the trained GCN model, when evaluated on the extracted subgraphs at each sparsity level.

For clarity, the abbreviations used in all figures are as follows: IG (Integrated Gradients), GBP (Guided Backpropagation), SAL (Saliency), and RF (RankFusion explainer).

As illustrated in Figure~\ref{fig:acc_compare}, the GEC-based subgraph extraction method consistently outperforms the TES approach across all explainers. The performance improvement is particularly notable in the case of GNNExplainer, where the GEC method achieves nearly double the accuracy of TES, when the subgraph retains less than 90\% of the original edges.

Moreover, regardless of the extraction method (TES or GEC), the ranking of explainers in terms of classification accuracy remains consistent. CaptumExplainer variants (specifically IG, GBP, and SAL) demonstrate the highest performance, followed by PGExplainer, with GNNExplainer yielding the lowest accuracy.

\label{sec:result}

\begin{figure*}[t]
    \centering
    \begin{minipage}{0.49\linewidth}
        \centering
        \includegraphics[width=\linewidth]{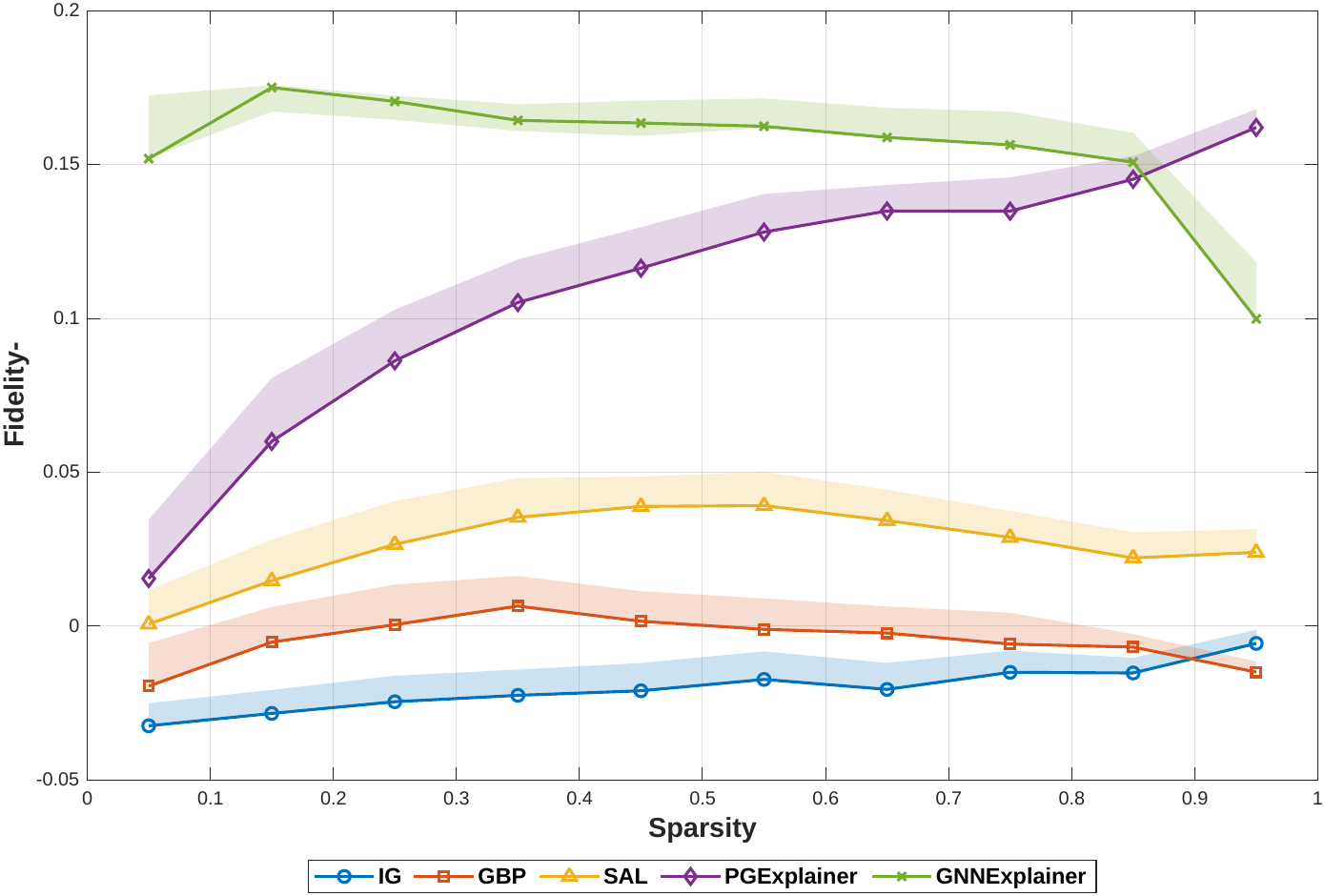}
    \end{minipage}
    \begin{minipage}{0.49\linewidth}
        \centering
        \includegraphics[width=\linewidth]{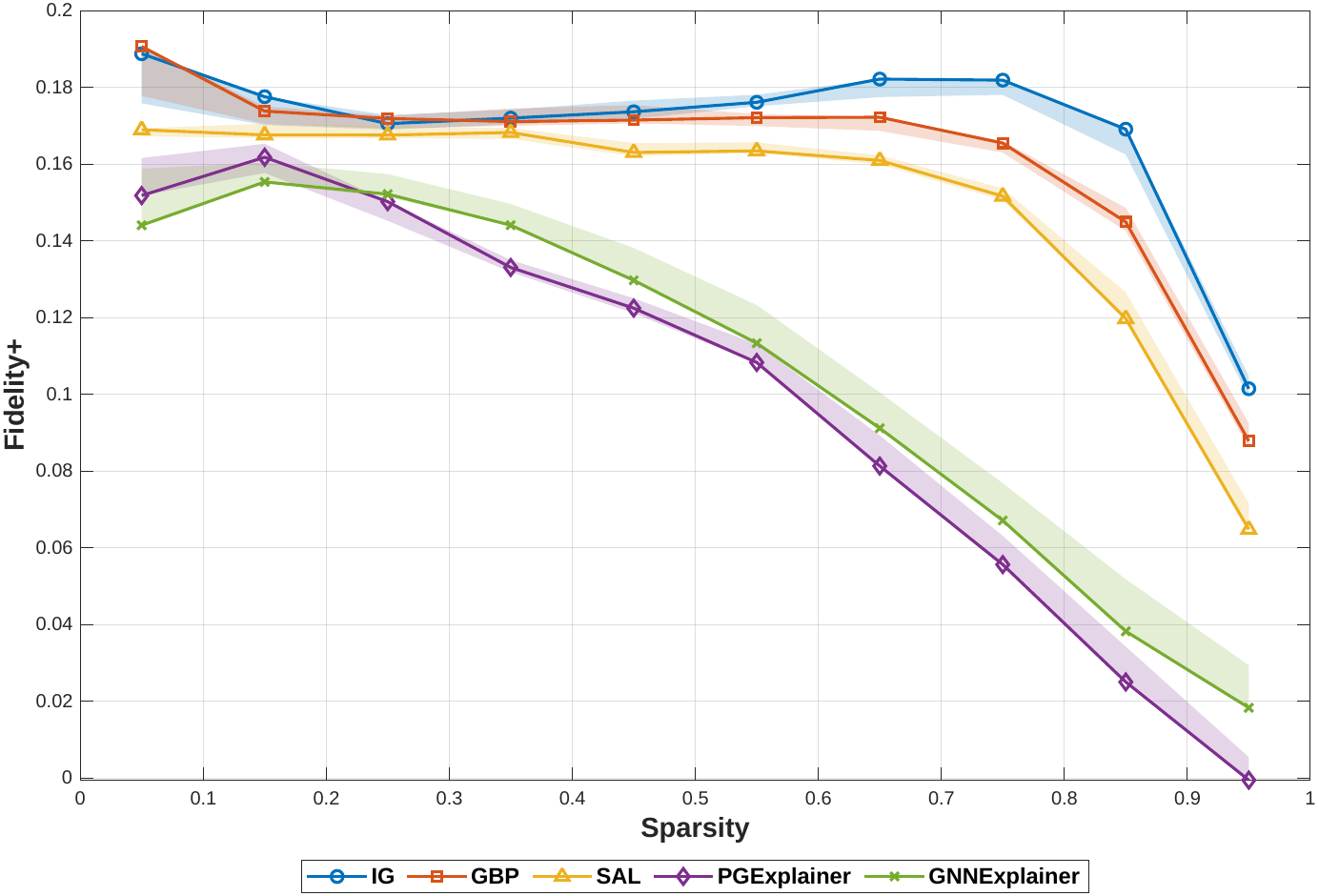}
    \end{minipage}
    \caption{Fidelity performance of explainability methods under the GEC approach.}
    \label{fig:fid_GEC}
\end{figure*}

We assess the performance of the explainers using the $Fidelity{-}$, $Fidelity{+}$, and Consistency metrics across different sparsity levels. Figure~\ref{fig:fid_GEC} presents the $Fidelity{-}$ and $Fidelity{+}$ results under the GEC approach. In each subplot, the x-axis represents the sparsity level, defined as the percentage of edges retained from the original graph, and the y-axis indicates the fidelity score. The solid lines show the fidelity values computed from the original, unperturbed graphs for each explainer and sparsity level.

To assess consistency in practice, we generated 20 perturbed versions of each input graph: 10 with random node removal and 10 with random edge removal. In our implementation, the perturbation ratio was set to \( p = 0.01 \).
Each perturbed graph was retained for analysis only if it preserved the model’s original prediction and the cosine similarity between the graph-level embeddings of the original and perturbed graphs was below a selected threshold. We performed a sweep over similarity thresholds \( \tau \) ranging from 0.01 to 0.03 in increments of 0.0025 to identify valid perturbations.
For each explainer, we visualized the variability of $Fidelity{+}$ and $Fidelity{-}$ scores across all valid perturbations. The solid lines represent fidelity values computed on the original (unperturbed) graphs, while the shaded regions reflect the range between the minimum and maximum fidelity scores observed across the valid perturbations.

The results show that $Fidelity{-}$ values are generally low and stable for explainers, when subgraphs are extracted using GEC. This is particularly evident for GNNExplainer, whose performance stabilizes significantly under this approach. According to the definition of $Fidelity{-}$, lower values are desirable, and this trend is clearly observed for the best-performing explainers (IG, GBP, and SAL), which consistently yield lower $Fidelity{-}$ scores.

In contrast, $Fidelity{+}$ values demonstrate a consistent pattern across sparsity levels. The top explainers (IG, GBP, and SAL) achieve higher scores, indicating a substantial change in model prediction when the important subgraph is removed. This suggests that the identified important subgraph indeed contains the critical predictive structure. Furthermore, the consistency bands in the $Fidelity{+}$ plot are generally narrow, indicating that explanations remain stable across perturbations. Although $Fidelity{-}$ bands are somewhat wider, the results still indicate reasonable stability, particularly for the top explainers. These findings highlight the advantage of using GEC in conjunction with effective explainers for generating reliable and robust explanations.

To assess the effectiveness of the proposed RankFusion explainer, we investigate whether selecting and combining the outputs of the top-performing explainers can lead to improved subgraph quality and model performance. Based on the previous analysis across accuracy, fidelity, and consistency metrics, the two most effective explainers identified were Integrated Gradients and Guided Backpropagation. RankFusion is applied by aggregating the edge importance rankings produced by these two explainers using the methodology described earlier.
To provide a more comprehensive evaluation, we also compare RankFusion against two additional aggregation baselines: mean aggregation and rank voting. For mean aggregation, the normalized edge importance scores of IG and GBP are averaged. For rank voting, three explainers (IG, GBP, and Saliency) are considered, and edges are selected based on a majority vote (i.e., 2 out of 3 agreement) among their top-k ranked edges.

\begin{figure}[h]
    \centering
    \includegraphics[width=0.45\linewidth]{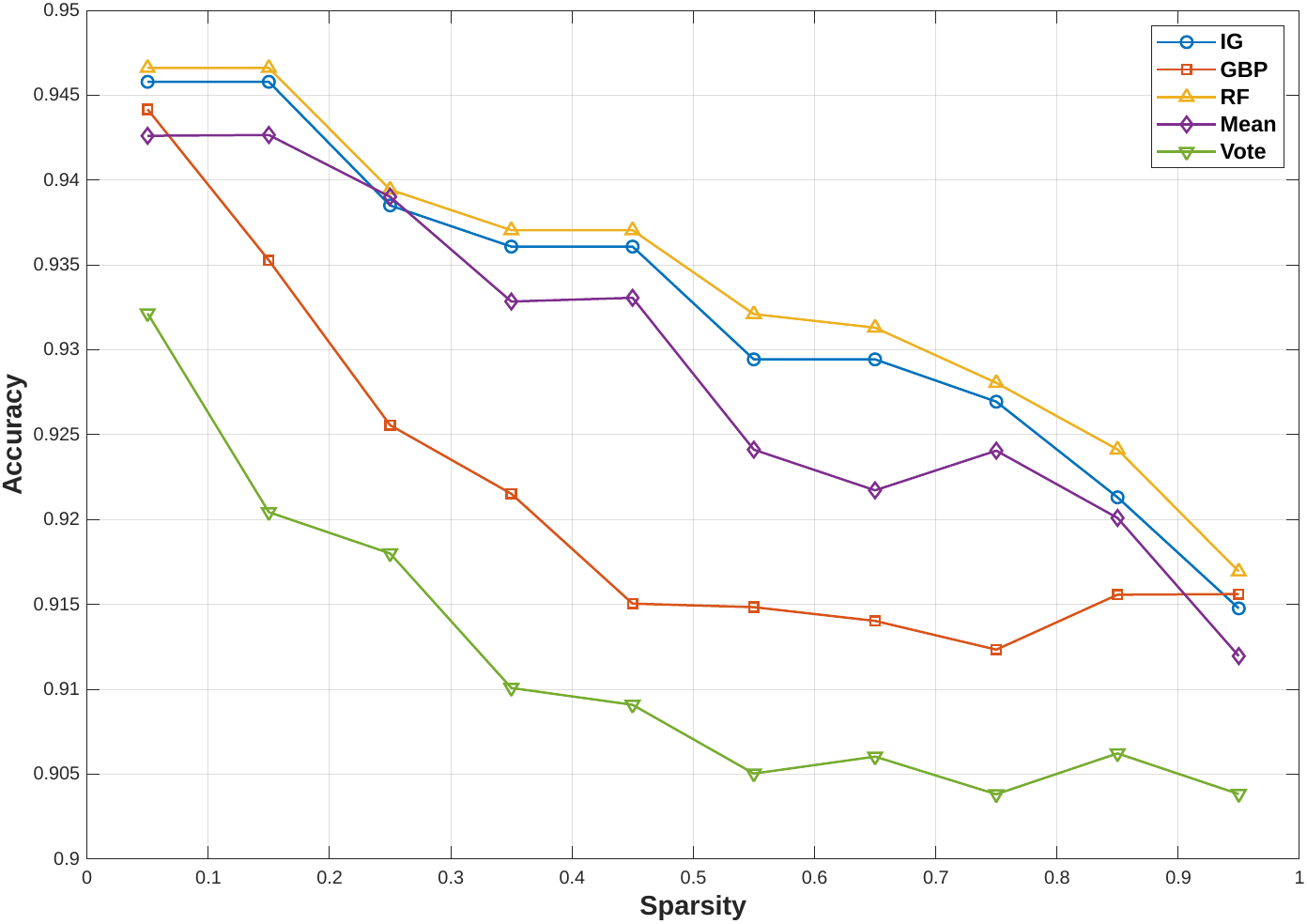}
    \caption{Accuracy performance of explainability methods, including individual top explainers (IG and GBP) and aggregation-based approaches (RankFusion, mean aggregation, and rank voting).}
    \label{fig:acc_agg}
\end{figure}

Figure~\ref{fig:acc_agg} compares the model's accuracy when using subgraphs generated by IG, GBP, and their aggregated result. All subgraphs in this evaluation are extracted using the GEC method. The x-axis represents the sparsity level, and the y-axis shows the accuracy obtained by feeding the resulting subgraph into the trained GCN. Across all sparsity levels, RankFusion consistently outperforms both the individual explainers and the alternative aggregation methods. While the improvement over the best individual explainer (IG) is modest, the results clearly support the hypothesis that selecting and combining the outputs of strong explainers leads to more informative subgraphs.

\begin{figure*}[h]
    \centering

    \begin{subfigure}[t]{0.32\textwidth}
        \includegraphics[width=\linewidth]{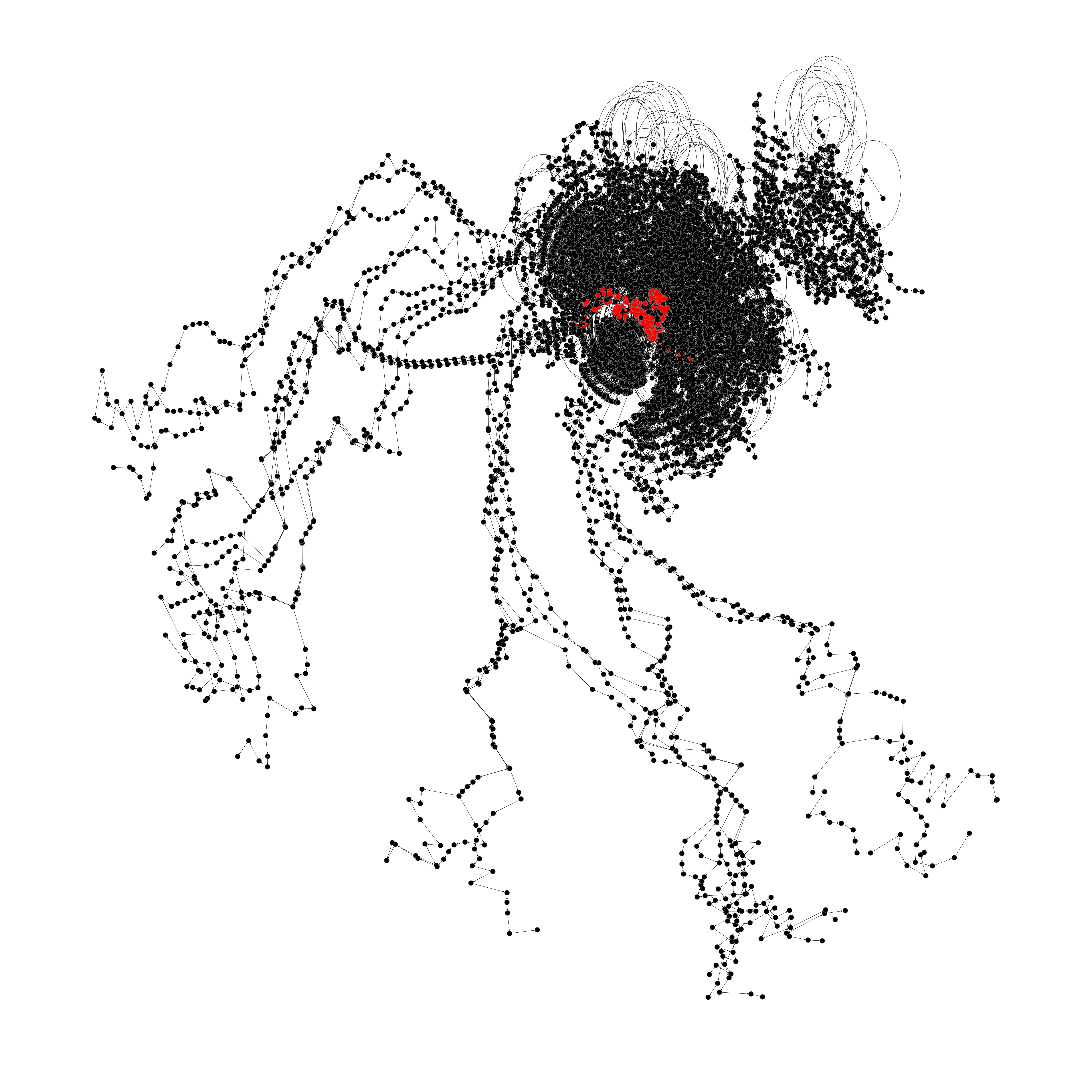}
        \caption{RankFusion explainer result.}
        \label{fig:sub1}
    \end{subfigure}
    \hfill
    \begin{subfigure}[t]{0.32\textwidth}
        \includegraphics[width=\linewidth]{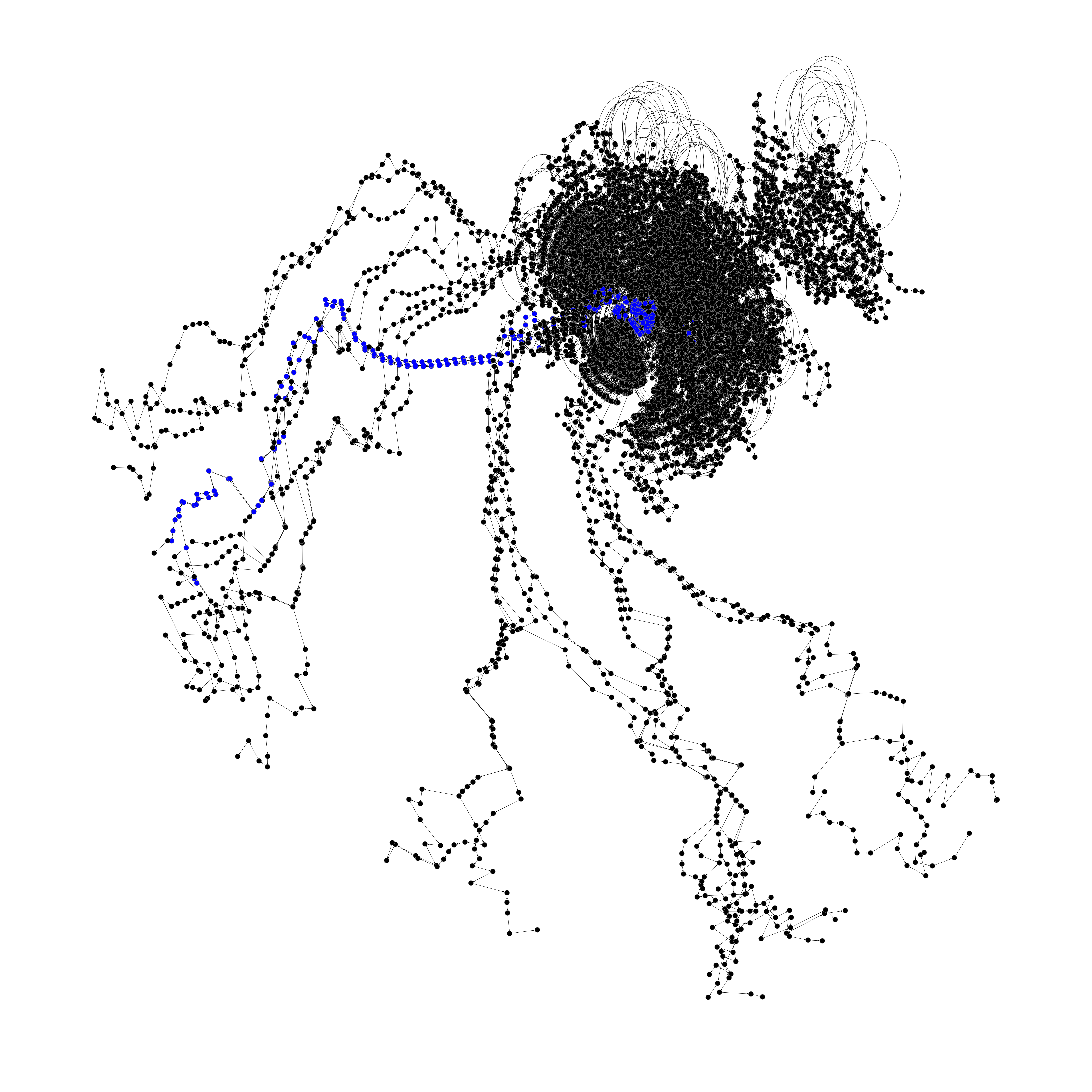}
        \caption{Integrated Gradients explainer result.}
        \label{fig:sub2}
    \end{subfigure}
    \hfill
    \begin{subfigure}[t]{0.32\textwidth}
        \includegraphics[width=\linewidth]{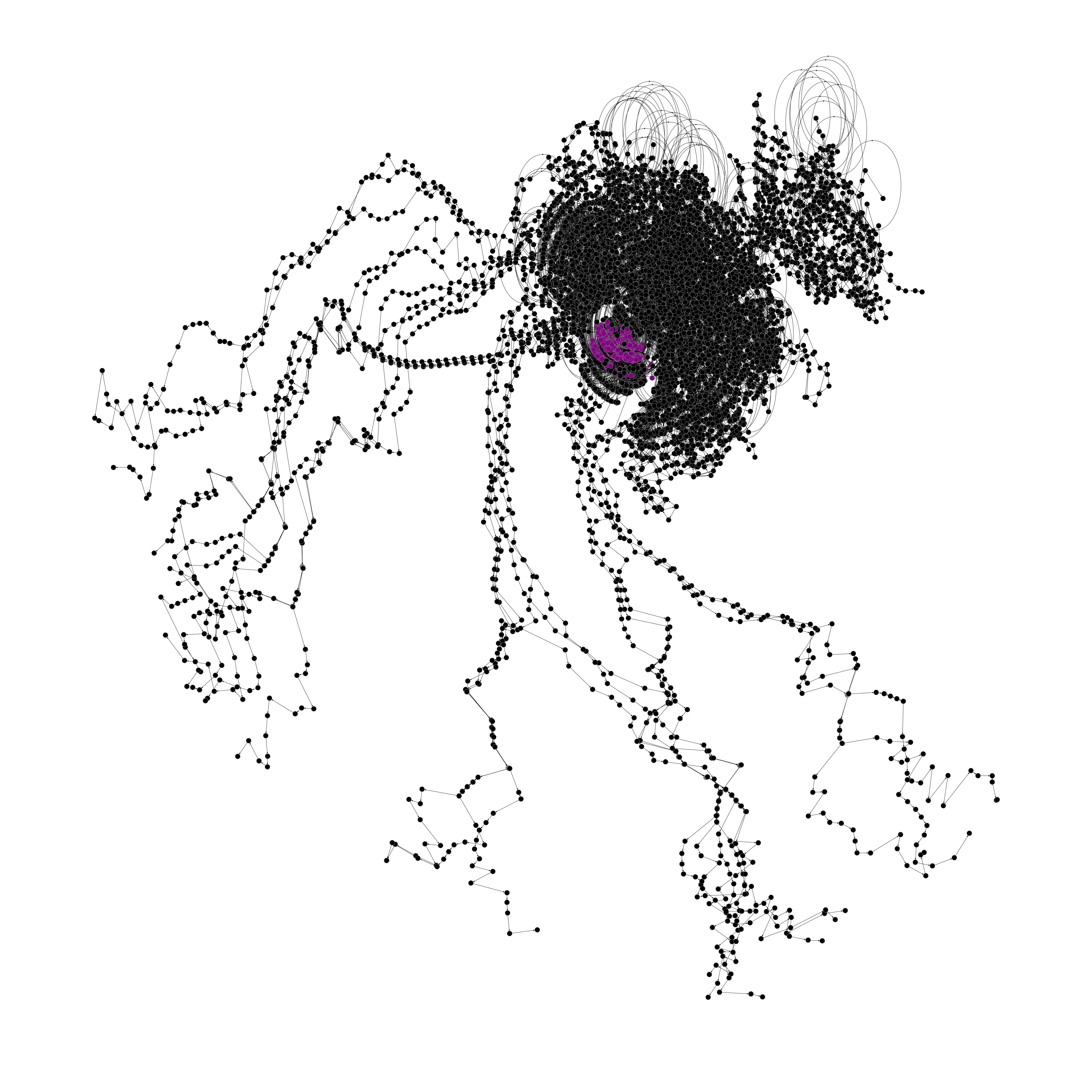}
        \caption{Guided Backpropagation explainer result.}
        \label{fig:sub3}
    \end{subfigure}

    \caption{Visual comparison of the top 5\% high-ranking subgraphs identified by three explainers on a malicious CFG.}
    \label{fig:main}
\end{figure*}

Figure~\ref{fig:main} visually compares the explanations produced by the RankFusion explainer, IG explainer, and GBP explainer on the same CFG of a malicious sample. In each subfigure, the colored region highlights the top 5\% of ranked edges along with their corresponding nodes, forming a connected subgraph that is considered most informative by the respective explainer.

\section{Conclusion}\label{sec:conclusion}
This paper presents a comprehensive framework for malware detection using dynamically generated CFGs. Node-level assembly instructions are embedded into a continuous vector space through an embedding approach that combines rule-based encoding with autoencoder-based learning. These representations are then used as input to a GNN-based classifier for malware detection.
To enhance the interpretability of the model, five explainers are employed: GNNExplainer, PGExplainer, and CaptumExplainer with three attribution methods, namely Integrated Gradients, Guided Backpropagation, and Saliency. In addition, we propose RankFusion, an aggregated explainer that selects and integrates the outputs of the top-performing explainers to improve the quality of generated explanations. We also introduce GEC as a connectivity-aware subgraph extraction strategy that mitigates the shortcomings of traditional top-edge selection.
Experimental results demonstrate the strong detection performance of the framework, with an accuracy exceeding 94\% and an F1 score above 96\%. Further analysis based on fidelity, accuracy, and consistency metrics shows that the combination of RankFusion and GEC produces more stable, interpretable, and accurate explanations, reinforcing the value of incorporating both explainer aggregation and structurally coherent subgraph extraction in GNN-based malware detection.
Future work will explore extending the framework beyond binary classification to identify specific malware families, enabling finer-grained threat analysis. Additionally, incorporating temporal features alongside CFGs could enrich the model’s understanding of malware dynamics. Another promising direction involves investigating the robustness of both the classifier and the explainers against adversarial perturbations, which is critical for maintaining reliability in security-sensitive applications.

\printcredits

\bibliographystyle{elsarticle-num}

\bibliography{cas-refs}



\end{document}